\documentclass{aa}
\usepackage{graphicx}

\begin{document}

   \title{Multiperiodicity, modulations and flip-flops in variable star light curves}
   \subtitle{I. Carrier fit method}
   \authorrunning{J. Pelt et al}
   \titlerunning{Carrier fit method}

   \author{J. Pelt
          \inst{1}
          \and
          N. Olspert
          \inst{1}
          \and
          M.J. Mantere
          \inst{2}
          \and
          I. Tuominen
          \inst{2,3}
          }

   \offprints{J. Pelt}

   \institute{
             Tartu Observatory, 61602 T\~{o}ravere, Estonia
             \and
             Department of Physics, Gustaf H\"allstr\"omin katu 2a (PO Box 64), FI-00014 University of Helsinki, Finland
             \and
             Deceased
             }

\date{Received --,--}

   \abstract
   {The light curves of variable stars are commonly described using
     simple trigonometric models, that make use of the assumption that
     the model parameters are constant in time. This assumption,
     however, is often violated, and consequently, time series 
     models with components that vary slowly in time are of great interest.}
   {In this paper we introduce a class of data analysis and visualization 
     methods which can be applied in many different
     contexts of variable star research, for example spotted stars,
     variables showing the Blazhko effect, and the spin-down of 
     rapid rotators.
     The methods proposed are of explorative type, and can be of 
     significant aid when performing a more thorough data analysis 
     and interpretation with a more conventional method.
   }
   {Our methods are based on a straightforward decomposition
     of the input time series into a fast ``clocking'' periodicity and
     smooth modulating curves. The fast frequency, referred to as the
     carrier frequency, can be obtained from earlier observations 
     (for instance in the case of photometric data the period can be
     obtained from independently measured radial velocities), 
     postulated using some simple
     physical principles (Keplerian rotation laws in accretion disks), or
     estimated from the data as a certain mean frequency. The
     smooth modulating curves are described by trigonometric
     polynomials or splines. The data approximation procedures are
     based on standard computational packages implementing simple or
     constrained least-squares fit -type algorithms.}
   {Using both artificially generated
     data sets and observed data, we demonstrate the utility of the
     proposed methods. Our interest is mainly focused on cases
     where multiperiodicity, trends or abrupt changes take place in
     the variable star light curves.}
  {The presented examples show that the proposed methods
     significantly enrich the traditional toolbox 
     for variable star researchers. 
     Applications of the methods to solve problems of
     astrophysical interest will be presented in the next papers of the
     series.}  

   \keywords{stars: variables: general - methods: data analysis}
   \maketitle

\section{Introduction}

The light curves of variable stars can display very different modes of
behaviour. They can be quite regular, often even periodic, or
relatively fast changes can be detected. 
The methods to deal with such observed data can roughly be divided
into two classes. The first class encompasses tools that are based on
Fourier
analysis methods, assuming that the light curve can be decomposed into
a set of periodic (harmonic) components. The components themselves can
then be interpreted as manifestations of, e.g., different pulsational
modes (as in asteroseismology), rotation, orbiting, and so forth. The
local behaviour of such multicomponent curves can be quite
complex. For instance, it is quite easy to build simple examples where
destructive interference between two nearby components with nearly
equal frequencies allows some third component to dominate the observed
curve. The detection of such local regularities often lead to the
usage of the second type of analysis tools, usually referred to as
the time-frequency analysis methods.
Instead of a global model, a certain set of local variability
elements is postulated, and then their change in time is followed.
Probably the best known method
of this kind is the wavelet transform.

For continuously observed signals the Fourier transform -based methods
and different types of time-frequency transforms are a well-developed
part of the traditional data processing work flow (see e.g. Poularikas
(ed.),~\cite{Poularikas}).  The situation is different in the case of
astronomical data. The irregular observing patterns and long time
gaps in the measured series significantly complicate the analysis 
(see e.g. the review by Schwarzenberg-Czerny, 2003, and references
therein or Chapter~5 in Aerts et al~\cite{Aerts}).  For the first type of methods, the computed frequency
spectra often show spurious peaks due to the regularities in the gap
structure of the input data.  The second class of methods, on the
other hand, can only be used for the short, more or less continuous,
subsets of data.

In this paper we propose a relatively new method, at least in the
astronomical context, to deal with irregularly spaced data. 
We start from the simple physical assumption that even if an object is
observed to exhibit irregular or semi-regular variability, some regular
process can be identified in the background.  
The best example of this is stellar rotation, normally quite stable
over long periods of time, the semi-regularity of the observed light
curve being caused by inhomogeneities, such as dark spots, on the
stellar surface.
Another relevant setting is provided by accretion disks, where the
Keplerian rotation of the accretion elements is more regular than the
observed outflow of radiation.
Using this assumption, we decompose the observed light curves into two
components: a rapidly changing carrier tracing the regular part of the
signal, and its slowly changing modulation. In this way, we construct
a method combining both aspects of the two classes of methods
presented above.

The proposed method is certainly {\em not} a new method to seek
hidden periods from data with gaps and irregularities.
Instead, we try to build
continuous models for irregularly observed data using {\it a priori}
known frequency. Proper visualization of the model can then help us
to interpret the data at hand and to reveal possible physical
effects. The method is of explorative nature and can be used as a
starting point before building more sophisticated statistical,
numerical or dynamical models.

The layout of the paper is somewhat non-traditional.
Instead of commencing with a lengthy introduction and comparison
of the new proposed method to various time series analysis methods
that have been used in variable star research, we start with
introducing the basic elements of the analysis method in
Sect.~\ref{CFmethod}, where we also give a detailed description of
the algorithms used. Sect. \ref{Examples} is dedicated to
demonstrate the suitability of the proposed method in various
astronomical settings, using both artificially generated and
observational data sets. This allows the reader to get
appreciation of the new method at work. In Sect.~\ref{RelatedWork}
we will give an overview of the related developments, discuss the new
method in the light of previous work, and in Sect.~\ref{Discussion}
discuss the domains of its applicability. 
The full-scale application
of the method to longer sets of observational data will be presented
in the forthcoming papers in the series.

\section{Carrier fit method} \label{CFmethod}

The notion of the carrier frequency is well known from communication
theory. 
Very often this is just a single harmonic tone with a fixed
frequency $\nu_0$. The carrier is modulated by a signal of interest
and transmitted to receiver. In the receiver the modulated signal is
demodulated and useful information is recovered. The simplest model
for such a process is just a waveform
\begin{equation}
f(t) = a(t)\cos (2\pi t\nu _0 ) + b(t)\sin (2\pi t\nu _0 ),\label{primi}
\end{equation}
where the carrier wave is modulated by two smooth low-frequency signal
components $a(t)$ and $b(t)$. 
In some cases the signal can be represented in an analytic complex form
\begin{equation}
f_a (t) = A(t)e^{2\pi j\varphi (t)}, 
\end{equation}
which allows the explicit tracking of the time dependent amplitude $A(t)$,
the phase $\varphi (t)$, and most importantly, the instantaneous frequency
\begin{equation}
\nu(t) = \frac{{d\varphi (t)}}{{dt}}.
\end{equation}
The real astronomical signals very often contain also higher harmonics
of the basic frequency.
Due to this we introduce, from the very beginning, a more complex
model of the form
\begin{equation}
f(t) = a_0(t)  + \sum\limits_{k = 1}^K {\big ( a_k (t) \cos (2\pi tk\nu_0 ) + b_k (t)\sin (2\pi tk\nu _0 )\big )}. \label{full}
\end{equation}
By including the term $a_0(t)$, we allow the mean level of the signal
to be different from zero.
We also assume that the total number of harmonics, $K$, is not very large.

\subsection{The algorithm}

To describe the details of the carrier fit method we start from the
simplest model of a variable star light curve:
\begin{equation}
f(t) = a(t)\cos ({2\pi t \over P _0 }) + b(t)\sin ({2\pi t \over P_0} ), \label{simplest}
\end{equation}
where $P_0$ is the period (measured in days) corresponding to the
carrier frequency $\nu_0=1/ P_0$ (measured in cycles per day), and the two
functions $a(t)$ and $b(t)$ describe slow changes of the curve. 
Here we define a slow process as one for which the characteristic
changes of the functions $a$ and $b$ occur on time-scales significantly
longer than the period of the carrier oscillation, $P_0$.

To build approximation curves for the observed data we firstly need to
have a proper value for the carrier period $P_0$, the determination of
which is discussed in detail in Sect.~\ref{CarrierFrequency}. Secondly,
certain analytical or numerical models for the both slow modulating
functions are required. These models themselves depend on certain sets of
parameters, as discussed in Sects.~\ref{Trig} and \ref{Spline} for the
chosen two representations. 
After reliably determining $P_0$ and formulating a suitable model for
the modulators, the remaining task is to find the model that best fits
the observations, and retrieve the corresponding parameters. This data
fitting process itself can be formulated as a standard 
least-square approximation procedure.
This procedure can, in general, involve both linear and non-linear
parameters. Because of the overall complexity of non-linear fitting, we
usually restrict our analysis to the linear regime.

Below we demonstrate the carrier fit tool using two different
methods for smooth approximation of the modulators: trigonometric
polynomials and splines. 
There are certainly many other possibilities - essentially every known
method of data smoothing from literature can also be used in this
context.
  
\subsection{Trigonometric approximation}\label{Trig}

We start from the relatively transparent method of trigonometric
approximation. This allows us to explicitly control the frequency
domain behaviour of our algorithm.

Let the time interval $[t_{min},t_{max}]$ be the full span of our input
data. Then we can introduce a certain period $D = C\times
(t_{max}-t_{min})$ 
for which the coverage factor $C$ is larger than unity
(typically $C=1.1-1.5$). Using the corresponding frequency, $\nu_{D}= 1/D$,
we can now build a trigonometric (truncated) series of the type:
\begin{equation}
a(t) = c_0^a  + \sum\limits_{l = 1}^L {\big ( c_l^a\cos (2\pi tl\nu _D ) + s_l^a\sin (2\pi tl\nu _D )\big ),} 
\end{equation}
and
\begin{equation}
b(t) = c_0^b  + \sum\limits_{l = 1}^L {\big ( c_l^b\cos (2\pi tl\nu _D ) + s_l^b\sin (2\pi tl\nu _D )\big ),} 
\end{equation}
to be used as models for smooth modulating curves. The coverage
factor $C$ must be larger than the data span to avoid artefacts due to the
built-in periodicity of our trigonometric models. 
Too large coverage factor leads to redundant parametrization. Very often the
building of the trigonometric model starts from the rounding of the full span period, $D$, and checking that the coverage factor
remains in reasonable limits.
According to our definition of a slow process, the period $D$
must be significantly longer than the carrier period $P_0$.

Using basic trigonometric identities it is now easy to show that the full
expansion of the carrier fit model will contain cosine and sine
terms of the form $\left\{ {\begin{array}{*{20}c}
      {\cos }  \\
      {\sin }  \\
      \end{array}} \right\} \big ( 2\pi t(\nu _0  \pm l\nu _D )\big )
$
where $l=0,\dots,L$.
Proper expansion coefficient estimates must be computed for every term in the
series for the fixed carrier frequency $\nu_0$ and ``data frequency''
$\nu_D$; this is a standard linear estimation procedure and can be implemented using standard
mathematical (statistical) packages.

In the beginning of our discussion about the algorithm, we started off
with the simplest light curve model, Eq.~(\ref{simplest}), including
only the lowest harmonic $k=1$. 
To generalise to the full case, Eq.~(\ref{full}), including also
higher harmonics, we will construct trigonometric representations
$a_k(t)$ and $b_k(t)$ for each overtone of the carrier $k\nu_0,\ 
k=1,\dots ,K$. The time dependence of $a_0(t)$ is also modelled
using special trigonometric polynomial. If all polynomials use the
same number of harmonics $L$ and we approximate separate cycles by
$K$-harmonic model, then the overall count of linear parameters to be
fitted is $N=(2 \times L+1)*(2 \times K+1)$.

The actual choice of the representative parameters $K$ and $L$
depends on the particular object we are working with. The number of
tones, $K$, depends on the complexity of the phase curves. The choice of
$L$ is constrained by the longest gaps in the time series. In
principle, it is also possible to compute certain formal constraints
using standard statistical regression theory (see Draper \&
Smith~\cite{Draper}). Very often the inspection of phase plots which
are computed with different parameter sets can also be useful.

\subsection{Spline approximation}\label{Spline}

The trigonometric models of the modulation curves are the most
useful when the actual behaviour of the modulations is
relatively continuous, i.e. without sharp transients. 
The spline approximations are more flexible from this point of view.

To build spline approximations we need to divide the full data time span
into, say $L$, intervals. For every interval with index $l=1,\dots,L$, and for
each overtone of the carrier $k=1,\dots,K$, we define two local sets of
cubic curves
\begin{equation}
a_{kl} (t) = \sum\limits_{m = 0}^3 {a_{klm} } t^m ,
\end{equation}
and
\begin{equation}
b_{kl} (t) = \sum\limits_{m = 0}^3 {b_{klm} } t^m ,
\end{equation}
to be used as local modulators for the different tones of the cosine and sine 
components of the carrier wave. The complete analytical model for the data will
then consist of cosine components
\begin{equation}
a_{klm} t^m \cos (2\pi k t\nu _0 ),
\end{equation}
and sine components
\begin{equation}
b_{klm} t^m \sin (2\pi k t\nu _0 ).
\end{equation}
There will altogether be $2\times 4 KL$ such components. To take
into account the general level of the signal, we need to include also the unit
component, after which the total number of modes to be fitted into data
will be $8KL+1$. 
The splines are not just a sample of local
polynomials, but must obey certain restrictions. 
In our simple case,
we demand that the approximation curves themselves, their first and
second derivatives should be continuous. As a result we need to
complement our set of fitting equations with $2\times 3 K(L-1)$ linear
constraints. The total number of free parameters will then be
$N=2KL+6K+1$. To perform the constrained least-squares fitting we used
routines from the AS 164 -package written by
Stirling~(\cite{Stirling}). The implemented algorithms are based on
the so-called Givens rotations (see Gentleman~\cite{Gentleman}) and allow
to solve weighted least-squares equations. The constraints are
effectively taken into account by assigning infinite weights to
the constraining equations.

\subsection{Construction of an analytic signal} \label{Hilbert}
One of the useful properties of the carrier fit method is that it
allows easily to compute Hilbert transforms for the model
waveforms. Using the Bedrosian theorem (Bedrosian~\cite{Bedrosian}), we can
easily compute
for term
\[
f(t) = a(t)\cos (2\pi t\nu _0 ) + b(t)\sin (2\pi t\nu _0 ),
\]
its Hilbert transform
\begin{equation}
h(t) = a(t)\sin (2\pi t\nu_0 ) - b(t)\cos (2\pi t\nu_0 ).
\end{equation}
By definition, the modulators $a(t)$ and $b(t)$ are smooth
functions, or in spectral terms, their spectra are concentrated around
the zero frequency. 
The carrier frequency is significantly higher, and therefore the
modulators can be removed from the transform. The complex analytic
signal (Gabor~\cite{Gabor}) can now be constructed as
\begin{equation}
w(t)=f(t)+jh(t),
\end{equation}
its amplitude as function of time
\begin{equation}
A(t) = \sqrt {f^2 (t) + h^2 (t)},
\end{equation}
and the instantaneous frequency
\begin{equation}
\nu (t) = \frac{{h'f - f'h}}{{2\pi (f^2  + h^2 )}}
\end{equation}
can be computed. For the details we refer the reader to the paper by
Vakman~(\cite{Vakman}).

Unfortunately the analytic signal approach is useful only for
waveforms which are either completely or nearly harmonic. For more
general curves the instantaneous values for amplitude, phase and
frequency can be formally computed, but their interpretation is not
straightforward.
 
\subsection{Estimation of the carrier frequency}\label{CarrierFrequency}

The most important input parameter for the carrier fit method is naturally
the carrier frequency $\nu_0$, or the corresponding period $P_0=1/\nu_0$
itself. Very often a useful value is already known from previous
data analysis steps. Here are some typical cases:
\begin{itemize}
\item Estimate from previous literature is at hand.
\item Orbiting period is known from eclipses.
\item Rotation period is computed from radial velocity data (say, in fully synchronized binaries).
\item The period is estimated by averaging its seasonal values (Blazhko effect stars, spotted stars).
\item The central frequency of the band in the Fourier spectrum can be used (Quasi-Periodic Objects). 
\end{itemize}
In most cases, therefore, an initial value can be easily determined.
The first trial can be improved later if the phase plots show significant
upward or downward trends.

In the case when we do not have a proper value for carrier frequency we
need to estimate it from the data.
Next we formulate a method to compare different
carrier frequencies, and describe how a rough initial value can be
improved to eliminate any trends in phase plots.  

The most natural choice to measure the quality of the
carrier frequency value is to inspect the stability of the resulting
light curve - with the optimal carrier frequency, the most stable
behaviour should be obtained.  For this purpose we can use the
interpolated carrier fit trial waveform.  Even if built using a very
rough preliminary estimate for the frequency $\nu_0$, it approximates
the input waveform in its entirety and is a continuous function. Let
us fix a certain trial frequency $\nu_T$ somewhere around the
preliminary carrier frequency $\nu_0$. We can divide the interpolated
waveform into intervals of the length $P_T=1/\nu_T$, and construct
local phase curves
\begin{equation}
\varphi _i (\phi ) = \hat f(t_i  + P_T \phi ),
\end{equation}
where $t_i = t_0+iP_T, i=0,1,\dots,N_T-1$ is a sequence of starting
points of the different intervals of the interpolated curve $\hat f(t)$,
phase $\phi$ runs from 0 to 1, and $N_T$ is the total number of the
intervals for a particular trial period. The phase curves change if
we move along time (later we will plot such time-dependent curves as
phase diagrams). We can now measure the rate of change of the phases
along time using a statistic
\begin{equation}
D^2 (\nu _T ) = \frac{1}{{N_T - 1}}\sum\limits_{i = 0}^{N_T  - 2 } 
{\int\limits_0^1 {(\varphi _i } (\phi ) - \varphi _{i + 1} (\phi ))^2 d\phi }. \label{D}
\end{equation}
Loosely speaking, we estimate the mean square of the first derivative
of the phase diagram along time. The optimal value for carrier frequency
should minimize our statistics (see the discussion in Sect.~\ref{carrier} and Fig.~\ref{Fg20}).

After computing the value for an optimal carrier frequency, we can
recompute the curve estimates and repeat the optimal frequency search,
and thereby obtain a refined value for the optimal carrier frequency.
In actual computations such a procedure takes only a
couple of iterations to converge.
Using this approach, uncertainties in the initial value of the carrier
frequency do not matter, as the method ultimately finds the optimal
and consistent carrier. 
However, one problem can be identified with this procedure. 
The optimal carrier frequency, as defined above, is essentially only a representative device.
Sometimes it may be hard to find a physical interpretation to to
the resulting carrier value computed from the data. Only if we can assume
that behind the seemingly erratic variation of oscillation parameters
there is a certain mean flow which can be characterized by a certain
mean frequency, we can accept the solution as physically
meaningful. Fortunately, in most practical cases, the carrier frequency
is available from the very beginning or physical considerations lead to
a proper interpretation easily.

\subsection{Visualisation}\label{visu}

We use the following scheme to visualize our results.  First we
calculate a continuous curve estimate, $\hat f(t)$, from the randomly
spaced and gapped data set, using a carrier-based least-squares
fitting scheme. This approximation is continuous and does not contain
gaps, and therefore allows us to get a smooth picture of the long-term
behaviour. Below we will discuss why and when we can trust such a
gap-filling scheme. Next we divide this continuous curve into segments
with a length of the carrier period $P_0 = {1 \over \nu_0}$. {\bf
  Throughout the paper, we measure periods in the units of days, and
  frequencies in the units of cycles per day.} We then normalize each
segment so that the approximating values span the standard range of
$[-1,1]$. After normalization, we stack the segments along the time
axis. To enhance the obtained plot, we somewhat extend every segment
along phases, so that the actual display is wider (along phases) than
a single period. The normalization is a relevant part of our procedure
because it helps to grasp phase information we are interested in
(trends, drifts, flip-flops etc). In the lower part of the plots we
often depict the distribution of the actual observational time moments
in the form of ``bar code''. The black stripes correspond to the
periods during which at least one observation is available. The bright
stripes correspond to gaps. This method of visualisation allows to
verify that the model fits into the data and not into the gaps. If the
number of nodes or harmonics $L$ for the modulation model curves is
chosen properly then the phase plots do not reveal underlying timing
structure.

\subsection{Local time-frequency analysis}

The obtained continuous, 'de-gapped' curve $\hat f(t)$ obtained by the
carrier fit method can be used as an input to different additional
analysis procedures. 
Now that a fully sampled data set has been obtained, the standard
approaches are applicable, due to which many different choices are
possible. For instance, we can compute local Fourier spectra for
sliding time windows, or compute various wavelet transforms (see
Poularikas (ed.), ~\cite{Poularikas} for additional possibilities). The
corresponding time-frequency plots can then be combined with phase
diagrams to get full grasp of the changes in the underlying processes
involved.

In particular, we use stacks of normalized, local in time, power
spectra to reveal trends in frequency structure of artificial or
observed curves.

\section{Examples and applications}\label{Examples}

The carrier fit method is quite straightforward in its formulation and
uses rather standard tools for its implementation. However, its full
utility can be appreciated only by looking at some actual
applications. To build up the required intuition, we first illustrate the
method applying it to very simple analytical light curve models which are sampled at
time points taken from real observations, and after
that, to real astronomical data sets. In the current context,
however, we totally ignore the physical background of the effects
observed in each example star - our goal here is to show the
versatility of the method.

\begin{figure}
  \resizebox{\hsize}{!}{\includegraphics{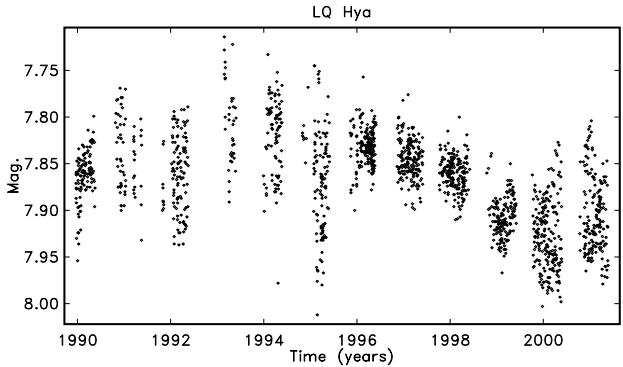}}
  \caption{V-band photometry of LQ Hya, JD 2447881 - 2452053. The longest time gap in the data set is 284 days.}
  \label{Fg01}
\end{figure}
\begin{figure}
  \resizebox{\hsize}{!}{\includegraphics{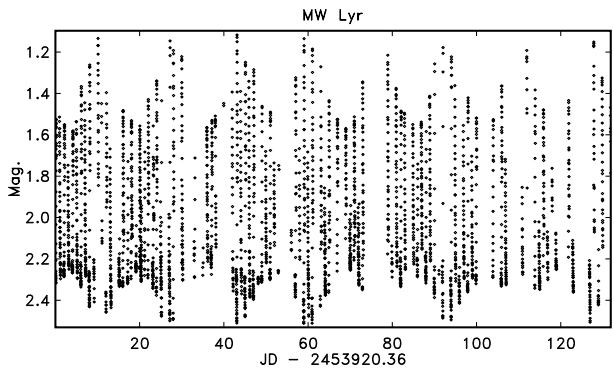}}
  \caption{V band photometry of MW Lyr, JD 2453920 - 2454052.}
  \label{Fg01a}
\end{figure}
\begin{figure}
  \resizebox{\hsize}{!}{\includegraphics{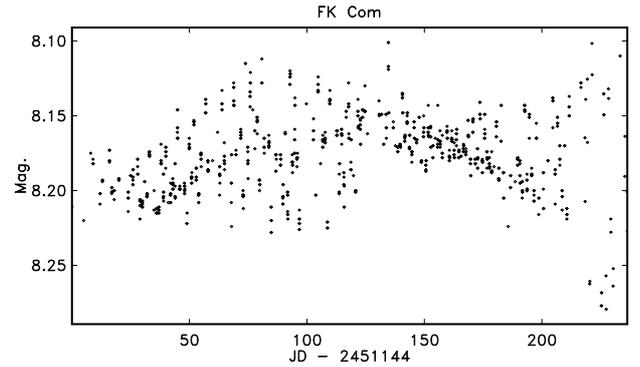}}
  \caption{V-band photometry of FK Com, JD 2451144 - 2451380. This is the only data fragment where the flip-flop phenomenon is observed continuously.}
  \label{Fg01b}
\end{figure}

\subsection{Timing sequences and data sets}\label{timing}

To get realistic long-term photometric timing sequences we use
a compilation of the V-band photometry for the variable star LQ Hya
(Berdyugina et al.~\cite{Berdyugina02}). The used subset of 1627 points
(displayed in Fig.~\ref{Fg01}) covers the interval JD 2447881-2452053 and
contains rather long gaps (up to 284 days). In the most of the
examples below we use only time points from this data set and
calculate artificial time-dependent process values using analytical
formulae. To model observational errors we add to each analytical
time series value a noise component $\varepsilon(t)$, which we generate as a
normally distributed random variable with standard deviation of 5\%
from the overall series amplitude. This level of scatter is quite
characteristic for reasonably good photometry,
corresponding roughly to the signal-to-noise ratio of 20.

To demonstrate the suitability of the new methods to analyse the so-called
Blazhko effect, we use the RR Lyr -type variable star MW Lyr. The data and
detailed photometric solutions for the extensive photometry can be
found in Jurcsik et al.~(\cite{Jurcsik08}). As an input for reanalysis we
used a well-sampled subset of the observations spanning over the interval JD
2453920 -2454052, containing 3216 points (see Fig.~\ref{Fg01a}).

The usability of the method in the context of the flip-flop phenomenon
(e.g. Jetsu et al.~\cite{Jetsu93}) is demonstrated using a subset of
the extensive photometry of the rapidly rotating variable star FK Com,
described in Korhonen et al.~(\cite{Korhonen07}) and depicted in
Fig.~\ref{Fg01b}. The
data set contains 968 points and spans over the time interval JD 2451144 -
2451380.

The periods used as analytical model parameters are also taken to
match some values that can be derived from observations; we return
to their actual deduction from the data in the next papers of the
series.

\subsection{Mismatched carrier frequency}\label{Mismatch}

\begin{figure}
\resizebox{\hsize}{!}{\includegraphics{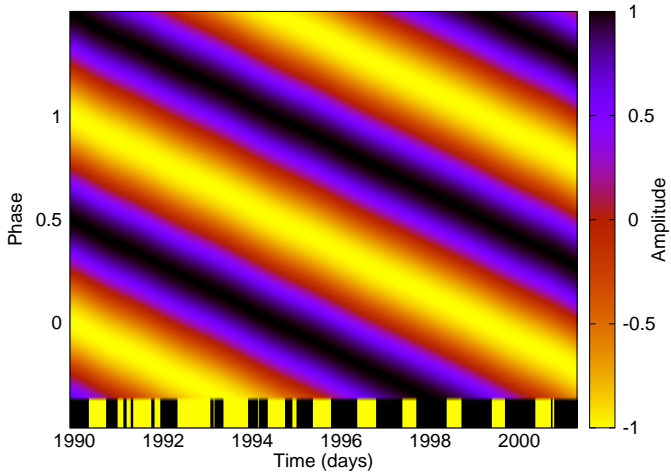}}

\caption{Time-dependent phase diagram, i.e. the light curve
amplitude profile over phase (y-axis) plotted as function of
time (x-axis), for the mismatched carrier frequency example
presented in Sect.~\ref{Mismatch}. In this plot, a slightly too
low carrier frequency of $\nu_0=0.148714$ is used, with $K=1$
model and $L=8$ modulation harmonics. Time points are obtained
from real V-band photometric observations of LQ Hya; the
bar-code in the bottom of the plot indicates when data has been
available (black) and the gaps (bright). For visualisation
details see Sect.~\ref{visu}.}
  \label{Fg02}
\end{figure}

Very often the appropriate carrier frequency $\nu_0$ is obtained from
information which is available before the actual data
processing. Being it rotation, orbiting or previously determined
frequency, we need to confirm or re-evaluate it. 
To do this, a model with smoothly varying coefficients is fitted into
the data, using the procedure described in
Sect.~\ref{CarrierFrequency}. If the actual frequency is stable, but
differs from the used carrier frequency we will see it from the
time-dependent phase diagram.  
In the following example we use artificially generated data, but use
the time-points from actual observations of LQ Hya described above.
For instance, if the input model is just a single harmonic
with a frequency $\nu=0.149009=1/6.711001$, magnitude offset and an added noise term $\varepsilon(t)$ (see Sect.~\ref{timing})
\begin{equation}
f(t)=12.0+\cos (2\pi \nu t)+\varepsilon(t),
\end{equation}  
then using a carrier value $\nu_0 = 0.148714 = 1/6.724333$, 
we will get a time-dependent phase diagram depicted in
Fig.~\ref{Fg02}. 
To stress that the actual time points used in the analysis are
taken from real observations with long gaps, we often plot a
stripe-like illustration of the observing times on the lower part of
the phase diagrams, as explained in detail in Sect.~\ref{visu}.
With the refined-search method for the carrier frequency, we can
recover the actual period in the input data set, after which the
tilted stripes in the figure become strictly horizontal.

\begin{figure}
\resizebox{\hsize}{!}{\includegraphics{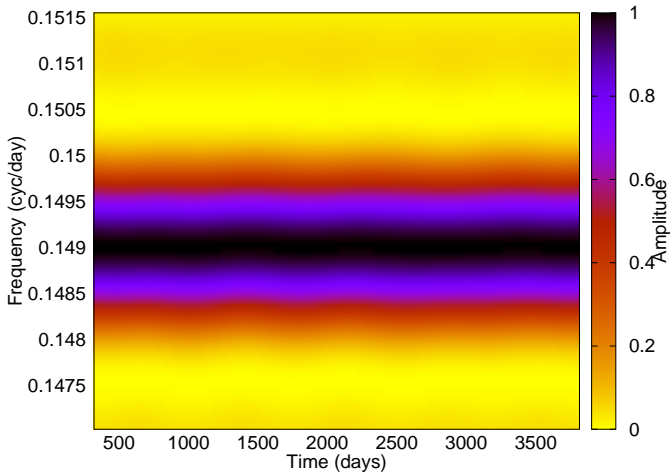}}
\caption{Time-dependent local spectra computed for the interpolated curve $\hat f(t)$, to recover the correct carrier frequency $\nu_0$. {\bf See Sect.~\ref{Mismatch}.}}
\label{Fg03}
\end{figure}

To analyse time-dependent frequency spectra, we use the same continuous
estimate, $\hat f(t)$, but now we use it as an input to a standard
time-frequency software\footnote{ISDA - Irregularly Spaced Data Analysis, available at http://www.aai.ee/$\sim$ pelt/soft.htm}. In our particular case we compute
frequency spectra for sliding time windows. After computing
each local in time spectrum we again normalize it to span the standard range
$[0,1]$. In this case the normalization helps us to track minor shifts
in frequencies even if the amplitudes of the separate components change
drastically. In this way our visualization method enhances the
aspects of the variability which we are looking for.
The result of this analysis is shown in Fig.~\ref{Fg03}.

\begin{figure}
\resizebox{\hsize}{!}{\includegraphics{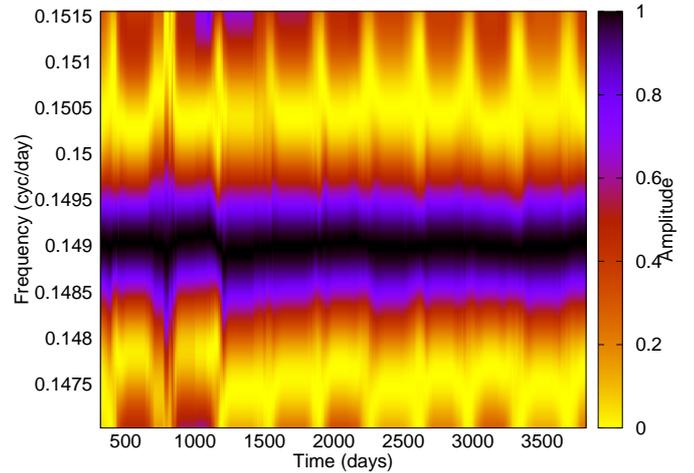}}
\caption{Time-dependent local spectra computed for the original data points without interpolation, to be compared with Fig.~\ref{Fg03}}
\label{Fg04}
\end{figure}

In principle it is possible to compute local frequency spectra for the
original (not interpolated) data set. In Fig.~\ref{Fg04} we depict the
results of this kind of transform. 
>From this plot it is evident how the gaps in the data tend to distort
the overall image. 
The changes in the spectra can be (mistakenly) counted as
real modulations.

\subsection{Abrupt change in the period}\label{Jump}
\begin{figure}
\resizebox{\hsize}{!}{\includegraphics{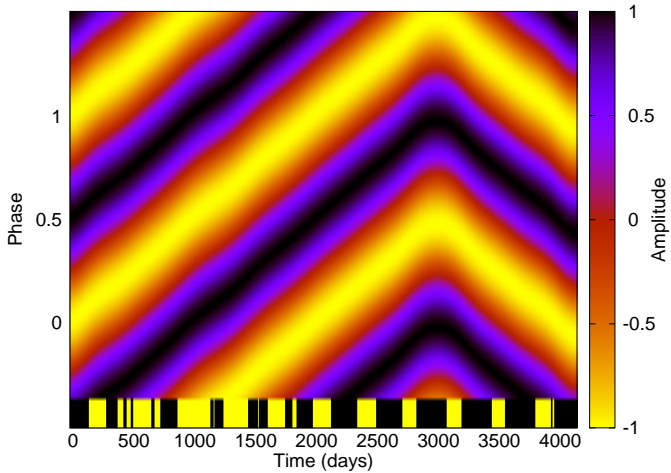}}
\caption{The effect of an abrupt change in the period
(Sect.~\ref{Jump}) in the time-dependent phase diagram. The
period before the jump is $P_1=6.747017$, and after it
$P_2=6.7018004$. The used carrier is $P_0=6.724333$, the number
of model harmonics, $K=1$, and the number of modulation harmonics,
$L=8$.}
\label{Fg05}
\end{figure}

\begin{figure}
\resizebox{\hsize}{!}{\includegraphics{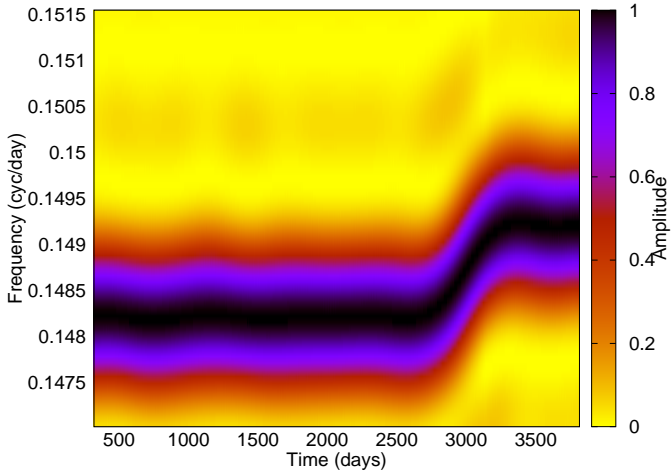}}
  \caption{The effect of an abrupt change in the period in the time-dependent local spectra; to be compared with Fig.~\ref{Fg05}.}
  \label{Fg06}
\end{figure}

The next interesting feature in variable star light curves is a
possible jump in the frequency. Applying the same time points from
observations as in the previous example, such a process can be
described with a simple time series model
\begin{equation}
f(t) = \left\{ {\begin{array}{*{20}c}
   {\cos (2\pi t /P_1) + \varepsilon (t),t < 3000,}  \\
   {\cos (2\pi t /P_2) + \varepsilon (t),t > 3000,}
\end{array}} \right.
\label{Jumpmodel}
\end{equation}
where $P_1$=6.747017, $P_2$=6.701800, $\varepsilon(t)$ is noise term
described above (Sect.~\ref{timing}) We choose the carrier frequency,
$\nu_0$, as a mean of the two model frequencies, and apply the carrier
fit method to build a continuous curve $\hat{f}(t)$; the results are
depicted in Fig.~\ref{Fg05}. As a comparison, local time-frequency
analysis is made, the results of which are plotted in Fig.~\ref{Fg06}.
The smooth appearance of the two plots is a result of the proper
approximation scheme.  The gaps of the input data are no longer
visible, as they are properly filled in. This facilitates the
visibility of the continuous patterns in the changing phase and
period.
However, if to compare the two plots, we see quite a relevant difference:
the transition time from one period to the other is much longer for
the time-dependent spectra if compared with the phase plot. To
explain the difference, we need to take into account the fact that the
time resolution of the transients differs for the two analysis
methods. The smoothness of the phase plots is determined by the
smoothness of the modulating coefficients $a(t)$ and $b(t)$, and
can be adjusted using different fitting schemes.  For time-dependent
spectra the smoothness is controlled by the width of the sliding
transform window. To obtain sufficient resolution in frequency domain
we lose resolution in time domain. This is a well-known and ubiquitous
uncertainty relation, and is common for all time-frequency analysis
methods.

\subsection{Two beating waves}\label{Beating}
\begin{figure}
\resizebox{\hsize}{!}{\includegraphics{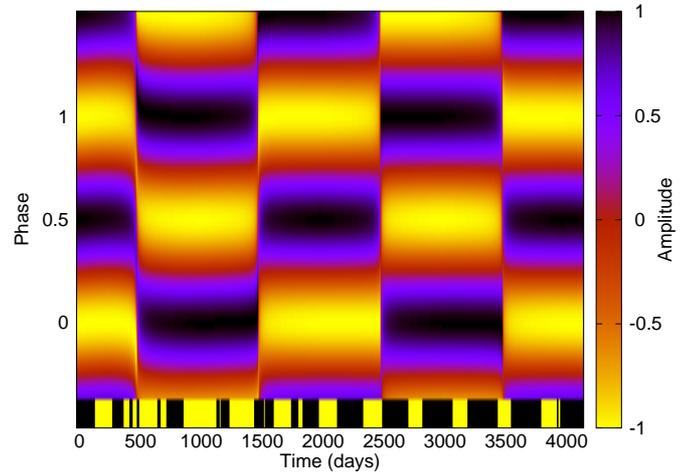}}
\caption{The effect of two beating periods, $P_1=6.747017$ and
  $P_2=6.701800$, in the time-dependent phase diagram. The carrier
  used is $P_0 = 6.724333$, the number of model harmonics $K=1$, and
  the number of modulation harmonics $L=8$. {\bf See
    Sect.~\ref{Beating}.}}
\label{Fg07}
\end{figure}

Next we consider an example, where an interference between two nearby
frequencies occurs. Again, the same time points with gaps as in the
previous examples are used, and an artificial time series model of the form
\begin{equation}
f(t) = \cos (2\pi t / P_1) + \cos (2\pi t / P_2) + \varepsilon(t), 
\end{equation}
where $P_1$=6.747017, $P_2$=6.701800, and $\varepsilon(t)$ is noise term (see Sect.~\ref{timing}),
is analysed using the carrier fit method to build a phase diagram,
plotted in Fig.~\ref{Fg07}, where a clear chessboard pattern can be seen.
This can be understood by considering the trigonometric identity
\begin{equation}
\cos (A) + \cos (B) = 2\cos (\frac{{A + B}}{2})\cos (\frac{{A - B}}{2}).
\end{equation}
The interference between the two nearby frequencies can be looked upon as
a certain high-frequency main tone modulated by a low-frequency beat
tone. In our case the beat period is $\approx 2000$ days. Because
we are dealing here with $100\%$ modulation, the sign of the beat
wave changes after every 1000 days, and the overall image of the
time-dependent phase looks like a chessboard. 
These abrupt changes in the phase are somewhat unexpected if we
take into account that the model contains only two simple harmonics.
The carrier fit method and the used visualization technique allows us
to reveal this natural flip-flop in the clearest way.

The continuous approximation curve $\hat f(t)$ can be further analysed
by using the local time-frequency analysis method.
\begin{figure}
\resizebox{\hsize}{!}{\includegraphics{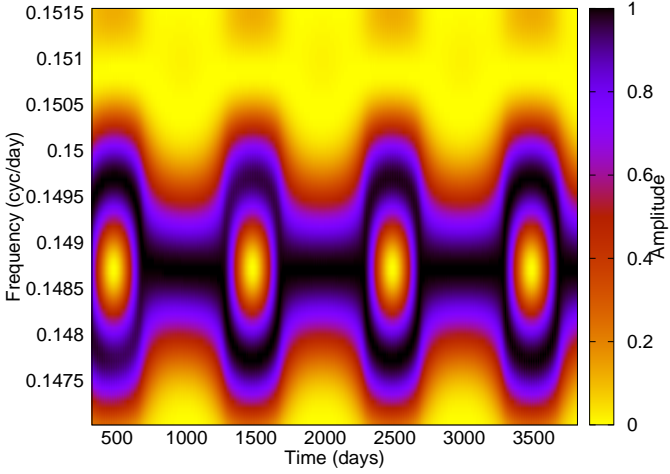}}
\caption{The effect of two beating periods in the time-dependent local spectra, sliding window length is 600 days; to be compared with Fig.~\ref{Fg07}.}
\label{Fg08}
\end{figure}
\begin{figure}
\resizebox{\hsize}{!}{\includegraphics{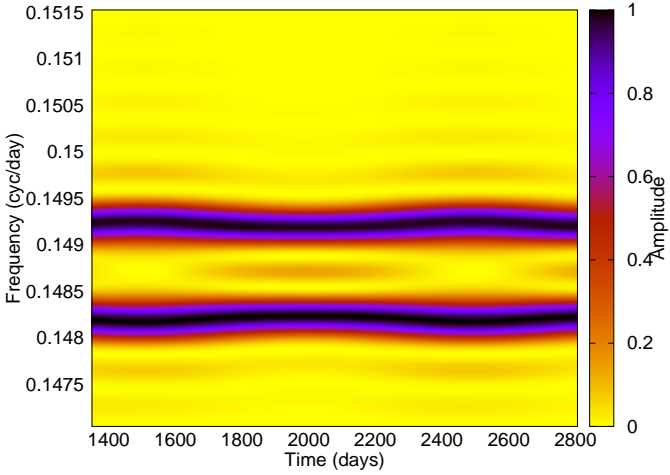}}
\caption{The effect of two beating periods in the time-dependent local spectra, sliding window length is 2400 days; to be compared with Figs.~\ref{Fg07} and \ref{Fg08}.}
\label{Fg09}
\end{figure}
The results of such an analysis with two different sliding windows
are shown in Figs.~\ref{Fg08} and \ref{Fg09}.
The first of these two figures, in which a sliding window of the
length 600 days was used, demonstrates a very serious problem of
the time-frequency analysis. From the form of the input
model, we expect that two parallel lines along time are
recovered. Instead, we see that the plot is dominated by a single maximum
at a period which corresponds to the mean of the two model
frequencies, and only around the locations, where the modulating beat
wave changes sign, the spectrum splits into two separate
frequencies. Even then the local maxima are not placed at the correct
periods. Such a weird picture is due to the usage of a relatively short
sliding time window (600 days) if to compare with beat period (2000
days). Even if we compute the local spectra for a four times longer time
window (as in Fig.~\ref{Fg09}), there is still a slow wave seen
in the positions of the two frequencies. 
This example illustrates a typical error made in time series analysis,
where the original data is divided into segments, and each segment is
analysed separately, after which conclusions are made according to the
obtained local periods. 
Were the local spectra computed without a preceding continuous
interpolation, the results could be even more misleading.  
In this particular case we consider the time-dependent phase plot a
more appropriate analysis and visualization method. 

\subsection{Linear trend in the frequency}\label{lineartrend}
As our next example of the carrier fit method, we define the input
data as follows
\begin{equation}
\label{changing}
f(t) = \cos (2\pi t((1 - \lambda(t) \times 0.148214 + \lambda(t)  \times 0.149214))+\varepsilon(t), \label{linmodel}
\end{equation}
where $\lambda(t)={t-t_{min} \over t_{max}-t_{min}}$ moves linearly from
$0$ to $1$ during the time span of the input data $[t_{min},t_{max}]$ and $\varepsilon(t)$ is a noise term (see Sect.~\ref{timing}).

\begin{figure}
\resizebox{\hsize}{!}{\includegraphics{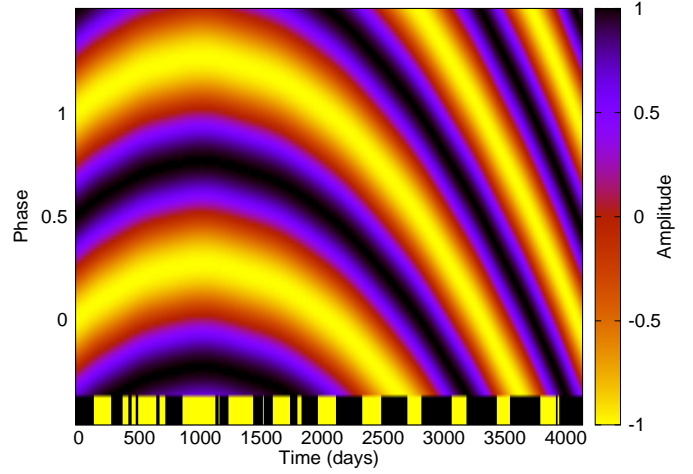}}
\caption{The effect of a linear trend in frequency, as
described in Sect.~\ref{lineartrend}, in the time-dependent
phase diagram. The used model parameters for the carrier fit
read $\nu_0=0.148714$, $K=1$, and $L=8$.}
\label{Fg10}
\end{figure}
\begin{figure}
\resizebox{\hsize}{!}{\includegraphics{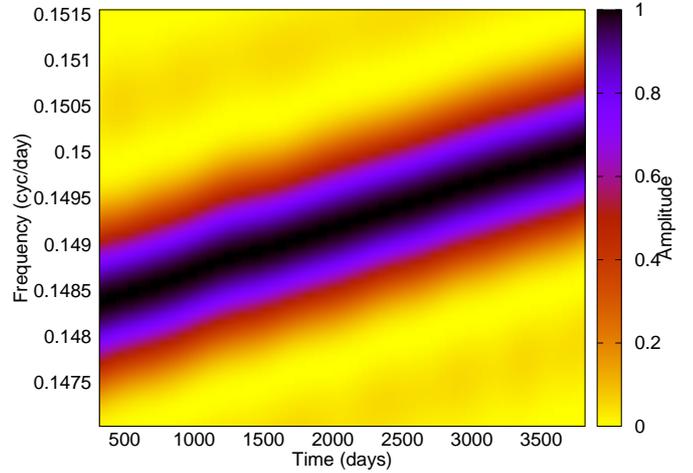}}
\caption{The effect of a linear trend in frequency in the
time-dependent local spectra. To be compared with
Fig.~\ref{Fg10}.}
\label{Fg11}
\end{figure} 

As seen from Fig.~\ref{Fg10}, the phase diagram for a curve with a changing
frequency is rather complex, 
and it is not easy to interpret the actual form of the variability in
the input data.
In this case, however, the local time-frequency analysis reveals quite
explicitly what is going on, as shown in Fig.~\ref{Fg11}.

\begin{figure}
\resizebox{\hsize}{!}{\includegraphics{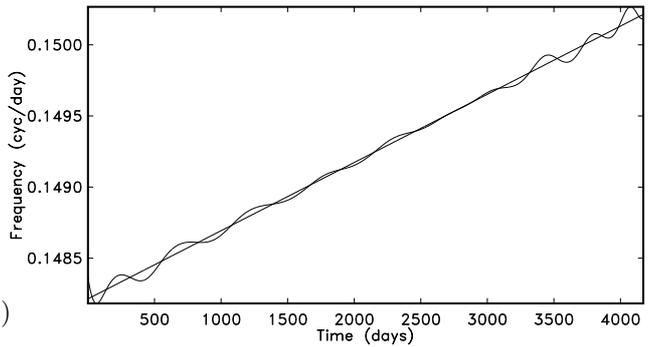}}
\caption{The instantaneous frequency estimated using the
Hilbert transform scheme of the carrier fit utility (thin line)
overplotted with the analytical model (thick line) for the example
presented in Sect.~\ref{lineartrend}. Parameters for the carrier
fit model: $\nu_0=0.148714$, $K=1$ and $L=8$.}
\label{Fg12}
\end{figure}

For this model, we can also apply an instantaneous frequency
estimation scheme based on the Hilbert transform to construct an
analytic signal, as described in Sect.~\ref{Hilbert}. 
In Fig.~\ref{Fg12}, the computed run of the instantaneous frequency is
displayed. Due to the noise component and gaps in the data set, the
recovery is not exact. 
Also, the boundary effects in the
beginning and end of the data are clearly visible,
resulting from the extra freedom for
the corresponding parts of the fitting curve. 
The results for this example also demonstrate an interesting aspect of
time-frequency analysis. 
>From Eq.(\ref{linmodel}) we can see that for the last time point, the
frequency in the input data is 0.149214. The instantaneous frequency
for this time point, recovered with both the local time-frequency
analysis and Hilbert transform methods, is somewhat higher than the
actual frequency.
The fact observed here, the input data frequency and the recovered
instantaneous frequency being not equal, is often overlooked (however,
see Hempelmann~\cite{Hempelmann}). The instantaneous frequency is not an
entirely local value, but it also depends on its near neighbourhood
due to the integral nature of the Hilbert transform.
 
\subsection{Nonharmonic components}\label{nonharm}

All the above examples were based on simple harmonic components. The
real variable star light curves are seldom so simple. To show how the
carrier fit method works for non-harmonic waveforms, we define a model
with two interfering components $f(t)=c_1(t)+c_2(t)+\varepsilon(t)$,
where both components are of the form
\begin{equation}
c(t) = \max (\cos (2\pi t\nu ),0),
\end{equation}
with the frequencies $\nu_1=0.148214$ and $\nu_2=0.149214$,
respectively, corresponding to the periods used in the example of two
beating waves, Sect.~\ref{Beating}. From this example, it is useful to
recall that the beat period of these two frequencies is roughly 2000
days.
In this model, however, the two waveforms are
``active'' only during a half of their cycles. 

It is quite clear that for
the modelling of the process we need higher carrier harmonics.
\begin{figure}
\resizebox{\hsize}{!}{\includegraphics{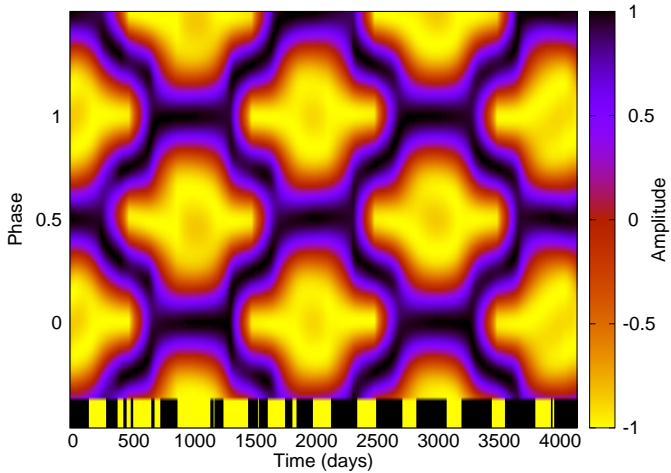}}
\caption{The effect of the interference of non-harmonic waves with
$\nu_1=0.148214$ and $\nu_2=0.149214$ in the time-dependent
phase diagram for a carrier fit model using $\nu_0=0.148714$. {\bf See Sect.~\ref{nonharm}.}}
\label{Fg13}
\end{figure}
In Fig.~\ref{Fg13} we show the results of the carrier fit with $K=3$.
Again, a regular pattern of minima and maxima is visible. If we
compare with the simple beat pattern of Fig.~\ref{Fg07}, the most
important difference is the occasional bi-modality of the phase
curves. This is a result of an interplay between phases of the two ``activity
zones''.  If the zones occur at phases nearby each other, we have
curves with a single maximum; in case of roughly 180$^{\circ}$
separation, two maxima appear. For more complex waveforms, the
interpretation of the phase plots can be even more convoluted.

\subsection{Blazhko effect}\label{Blazhko}

As an example of an application of the carrier fit algorithm, based on 
trigonometric modulation curves, to real observational data, we 
re-analyse a subset of the photometry for the RR Lyr-type star MW
Lyr, collected by Jurcsik et al. ~(\cite{Jurcsik08}). This pulsating
star is known to exhibit a modulation of its light curve in addition
to the variability due to the pulsations; this effect is known as the
Blazhko effect. Jurcsik et al.~(\cite{Jurcsik08}) proposed a bi-periodic
model with the very accurate pulsational period of $P_0$=0.397680
days modulated by a significantly longer period of $P_b$=16.5462 days.

\begin{figure}
  \resizebox{\hsize}{!}{\includegraphics{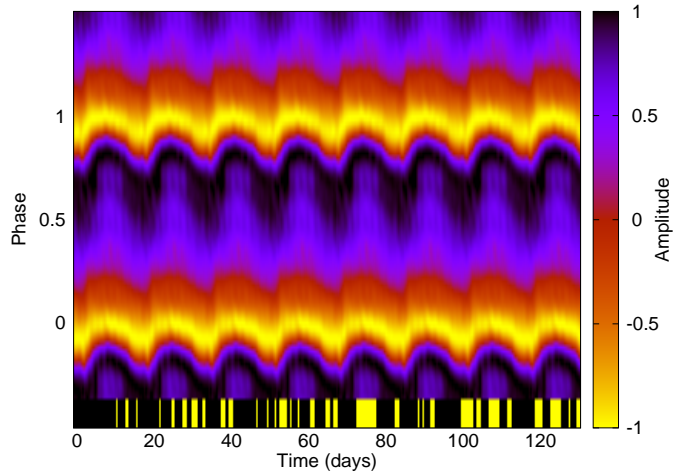}}
  \caption{Blazhko effect, phase diagram, bi-periodic fit with $P_0=0.397680$ ($K=5$), $P_b=16.5462$ ($L=10$). {\bf See Sect.~\ref{Blazhko}.}}
  \label{Fg14}
\end{figure}

We start from a similar bi-periodic model with the frequencies
$k\nu_0+l\nu_b$, where $k=1,\dots,5$ and $l=-10,\dots,10$, and
$\nu_0=1/P_0$ and $\nu_b=1/P_b$. The main difference to the original
analysis by Jurcsik et al.~(\cite{Jurcsik08}) is the usage of a smoother
form of the basic waveforms ($|k|\le 5$ instead of $|k|\le 13$), and a more detailed modulation model ($|l|\le 10$ instead of $|l|\le 4$).
The resulting normalized phase diagram
is presented in Fig.~\ref{Fg14}.
\begin{figure}
\resizebox{\hsize}{!}{\includegraphics{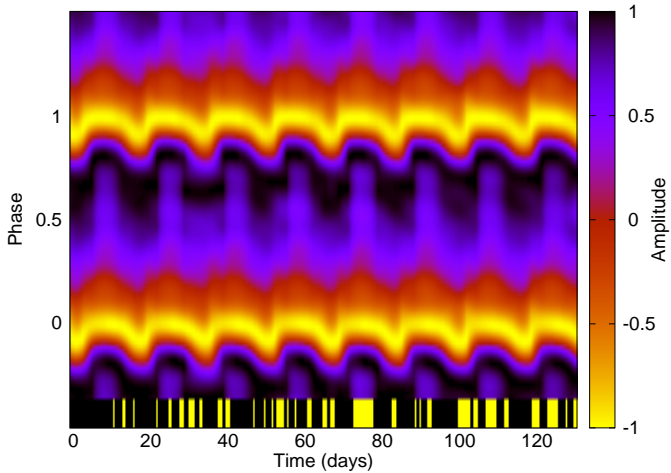}}
\caption{Blazhko effect, phase diagram, carrier fit with $P_0=0.397680$, $D$=150, $K=5$, and $L=10$. {\bf To be compared with Fig.~\ref{Fg14}.}}
\label{Fg15}
\end{figure}

To apply the carrier fit method to the same input data sequence, we fixed the
``data period'', D, to be $150$ days with the coverage factor of
$C\approx 1.14$, so that the trigonometric modulation curves cover
slightly more than the full data span. It is important to state that
the resulting phase plot 
is not very sensitive to this
number.  We also set the number of harmonics $K=5$ for the base
frequency, and $L=10$ for the modulations.  In this way the system of
linear equations to be fitted into the data was equal to the number of
equations in the bi-periodic fit. The results are presented in a phase
diagram of Fig.~\ref{Fg15}, showing a striking similarity of the phase patterns to the ones
plotted in Fig.~\ref{Fg14}.  An important point here is that the
characteristic modulation pattern is revealed without pre-set (or
precomputed) value for the modulation period. The only difference
between the two diagrams is the somewhat larger number of small
details for the bi-periodic fit and its exact periodicity. The phase
diagram for the carrier fit solution is somewhat smoother and reveals
minor fluctuations in the general periodic pattern.  Whether these are
fluctuations resulting form the observational noise, method artefacts,
or sampling irregularities, remains currently open.

As this example clearly demonstrates, it is possible to recover the
overall bi-periodic phase structure even without knowing the value of
one of the periods.

\subsection{Flip-flop}\label{flipflop}

To demonstrate the spline method in action, we performed
both a trigonometric and spline-based fit into a subset of photometry
of the star FK Com. A flip-flop phenomenon has earlier been detected
in this very subset of data (Jetsu et al.~\cite{Jetsu93}) covering one single
observing season, during which a sharp jump of roughly 0.5 in phases
occurs.
\begin{figure}
\resizebox{\hsize}{!}{\includegraphics{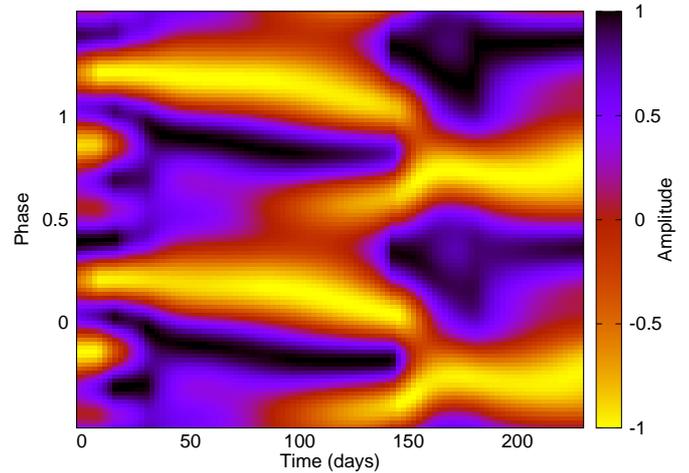}}
\caption{FK Com flip-flop, phase diagram, carrier fit with trigonometric modulations, carrier $\nu_0=2.40025$, $K=3$, $L=3$, $N=49$. {\bf See Sect.~\ref{flipflop}.}}
\label{Fg16}
\end{figure}
\begin{figure}
\resizebox{\hsize}{!}{\includegraphics{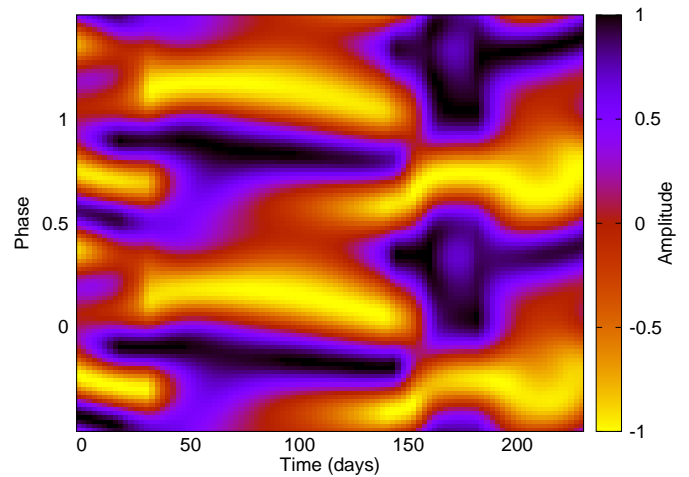}}
\caption{FK Com flip-flop, phase diagram, carrier fit with spline modulations, carrier $\nu_0=2.40025$, $K=3$, $L=5$, $N=49$. {\bf To be compared with Fig.~\ref{Fg16}.}}
\label{Fg17}
\end{figure}
The results of our analysis are shown in Figs.~\ref{Fg16} and
\ref{Fg17}; for both the trigonometric and spline fits, the general
picture is quite similar, and the abrupt jump in phase is
clearly revealed.
There are, however, some subtle differences. 
The trigonometric method is in a sense
more ``global'', due to which sharp features in one location can show up
as oscillations somewhere far away. The spline method is essentially
more ``local'', and as a result the picture is somewhat smoother. We
deliberately chose our parametrization here so that the total number of
free parameters is equal for both methods.

\subsection{Computation of the carrier frequency}\label{carrier}

\begin{figure}
\resizebox{\hsize}{!}{\includegraphics{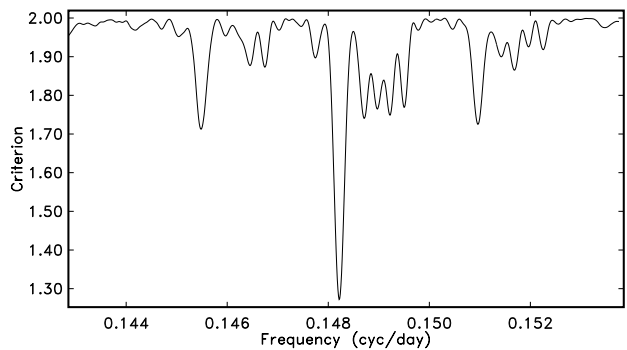}}
\caption{The Stellingwerf's statistic (criterion) for the input
data with an abrupt change in the period {\bf (the same model as used in Sect.~\ref{Jump})}, the strongest minimum is
at $P_0 = 6.7470$. Note the overall complexity of the spectrum. {\bf See the discussion in Sect.~\ref{carrier}.}}
\label{Fg18}
\end{figure}

\begin{figure}
\resizebox{\hsize}{!}{\includegraphics{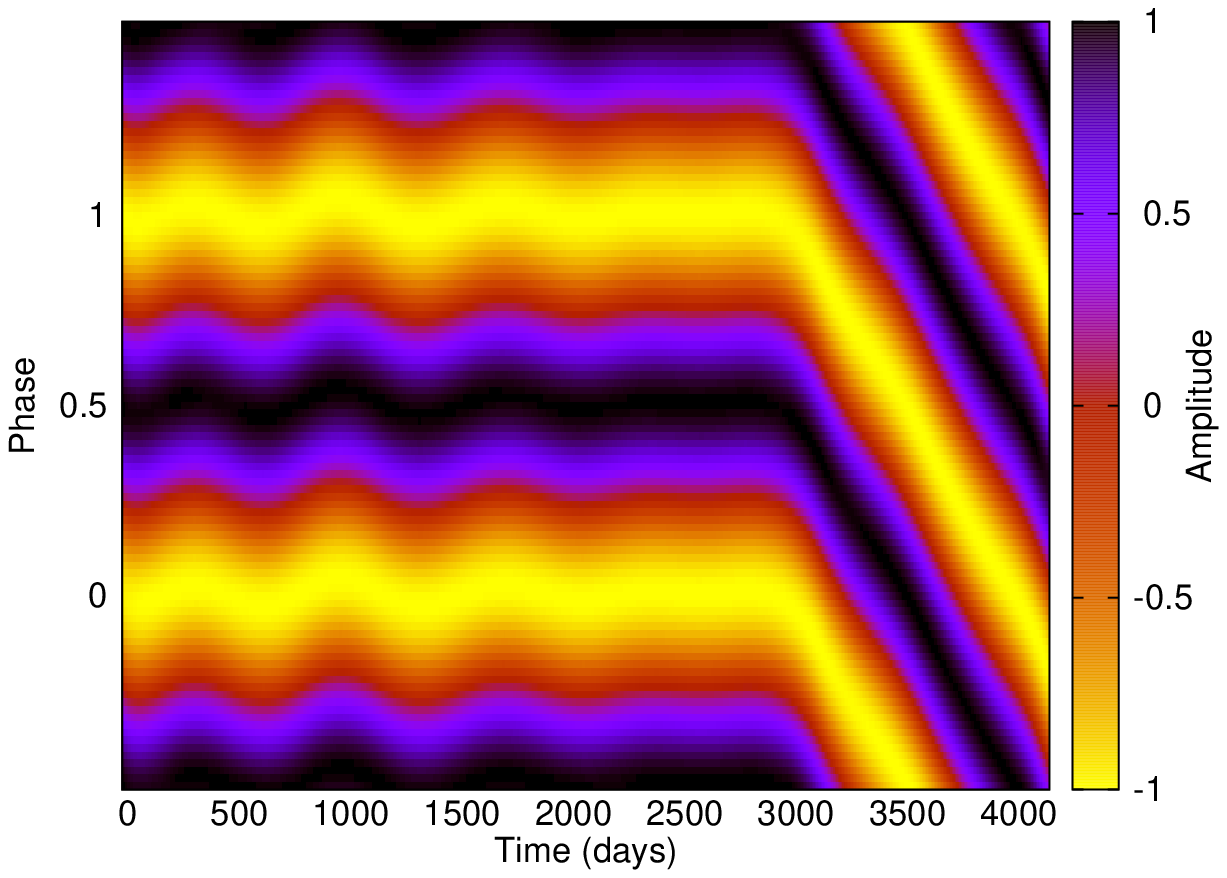}}
\caption{Abrupt change in the period, phase diagram, $P_1=6.747017$, $P_2=6.7018004$, carrier $P_0=6.7470$, $K=1$, $L=8$. {\bf See the discussion in Sect.~\ref{carrier}.}}
\label{Fg19}
\end{figure}

\begin{figure}
\resizebox{\hsize}{!}{\includegraphics{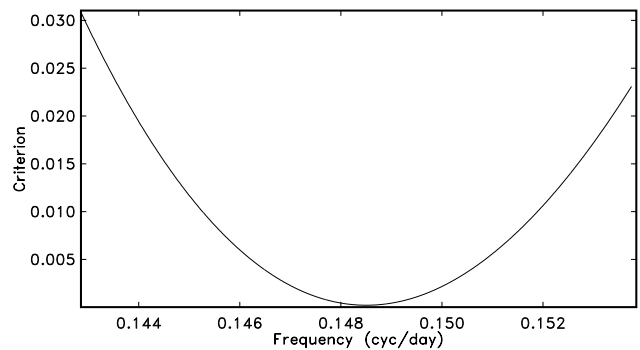}}
\caption{Phase-change spectrum, statistic $D^2$ (see Eq.~\ref{D}) is used as the criterion to estimate stationarity of the phase flow. {\bf See the discussion in Sect.~\ref{carrier}.}}
\label{Fg20}
\end{figure}

\begin{figure}
\resizebox{\hsize}{!}{\includegraphics{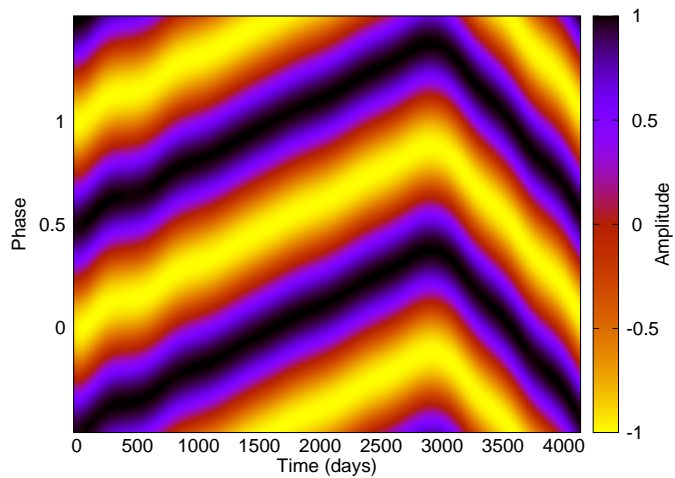}}
\caption{Abrupt change in the period, phase diagrams, $P_1=6.747017$,$P_2=6.7018004$, carrier $P_0=6.7334$, $K=1$,$L=8$. {\bf See the discussion in Sect.~\ref{carrier}.}}
\label{Fg21}
\end{figure}

Finally, we illustrate the method to estimate the carrier frequency. 
Our input data model is the same as used in Sect.~\ref{Jump},
Eq.~(\ref{Jumpmodel}), containing an abrupt change in the period.
The noise added to the input data through the term $\varepsilon(t)$ is
now increased to 20\% level.
A fragment of a standard phase-dispersion spectrum
(Stellingwerf~\cite{Stellingwerf}) for this data set is depicted in
Fig.~\ref{Fg18}. The strongest peak is at a period $P_0 = 6.7470$,
coinciding closely with $P_1$ in the input data model. This is to be
expected, as the input data consists of a long fragment with this
particular period. We proceed by 
using it as a carrier period value to build a time-dependent phase
diagram depicted in Fig.~\ref{Fg19}. 
This carrier value produces a nearly horizontal stripe with some
oscillatory behaviour for the part of the data having a period close
to the used carrier, and a strongly tilted pattern for the part with a
shorter period, already familiar from Sect.~\ref{Mismatch} for the
case of a mismatched carrier frequency. The situation can be improved
by refining the carrier frequency with a procedure described in
Sect.~\ref{CarrierFrequency} using the statistic $D^2$ defined by
Eq.~(\ref{D}), measuring the speed of changes in the phase
diagram. For this example calculation, we plot this statistics in
Fig.~\ref{Fg20}, showing a well-defined minimum at the frequency
$\nu_0 = 0.14850 = 1/6.7334$. Using this frequency as the carrier, a
time-dependent phase plot with minimum changes can be obtained, and is
depicted in Fig.~\ref{Fg21}.
Now the decreasing and increasing fragments of the phase diagram are
well balanced. 

Here we again note that the optimum carrier value is just a formal
construct, and not based on any physical principle or
insight. However, the procedure can be used as the last resort if no
{\it a priori} information is available.


\section{Related work} \label{RelatedWork}

Next we discuss the proposed algorithms in the
context of the other well-known analysis methods for variable star light curves.

\subsection{The O-C method}

In the classical O-C (observed versus computed) method (see for
instance Sterken, ~\cite{Sterken}), the input light curve is divided into
short fragments, so that for each fragment it is possible to securely
determine a certain light curve event, most often a local minimum or
maximum. 
For each event, a difference between the observed (O) and the computed (C) time moments is computed. The computed values are
commonly just values from a periodic sequence
\begin{equation}
T_E = T_0+PE,
\end{equation}
where $P$ is a period estimated beforehand, and $E$ is an integer count of
the full phases from the starting epoch $T_0$. The differences $O-C$
are then plotted for a full time span of the input data, and a proper
model for the deviations is then sought for. Most often polynomial
models are used.

In the carrier fit method proposed above, we are not very far off from
this traditional scheme. The carrier period, $P_0=\nu_0^{-1}$, is
just an analogue of the period $P$ in the equation for computed time point
values. The fitting procedure itself tries to tightly model local
fragments. In doing so, the local phase properties of the carrier wave
indicate peculiarities of the original curve (minima and maxima among
others). And finally, the smooth modulation curves are just analogues
of the traditional polynomial models for $O-C$ differences. In
principle, it is even possible to modify the carrier fit method so that
instead of trigonometric or spline models, simple polynomials are
used as modulation curves.

The $O-C$ method is often used to refine the already roughly
estimated periods. As shown above, a mismatch in the carrier
frequency reveals itself as a linear trend in the phase diagram.
Therefore, to estimate the needed correction to the carrier period, the
linear model for carrier modulations is considered sufficient.

The carrier fit method even shares the well-known problems of the
$O-C$ method. Most importantly - the wrong (or not exact enough) base
period $P$ can totally spoil the $O-C$ analysis. This is also true
when the choice of the proper carrier frequency is considered.
     
\subsection{Fourier analysis}

Methods based on, or connected to, standard Fourier analysis are
certainly the most popular in variable star research. The classical series
of papers, starting from Barning~(\cite{Barning}), and ending with
Koen~(\cite{Koen}) and Frescura et al.~(\cite{Frescura}), gives
a statistically sound treatment for the case when the input data can be
described by a simple harmonic. The generalizations for non-harmonic
(but periodic) and multiperiodic cases are obvious (see the chapter
``Frequency Analysis'' and references in the monograph by Aerts et
al.~\cite{Aerts}).  For us the important point is that the classical
methods are based on the assumption that the Fourier spectra of the
underlying processes are discrete, and form a sparse set of
frequencies. The sparseness of the spectrum makes it possible to recover
the constituent waveforms from data sets with long
gaps. Unfortunately, the periodicities caused by the gaps
can be a source of serious problems. Along with every discrete peak
in the original spectrum there will be a set of spurious peaks, and
a perfect recovery can be a challenge.

Another method of frequency analysis is based on the idea of
inversion. In this type of methods, the full discretized Fourier
spectrum is fitted into data (see for instance Kuhn~\cite{Kuhn} or
more recently Nygr\'en \& Ulich~\cite{Nygren}). For a sensible
inversion, the input data needs to be sampled quite densely. 
It must also be noted that inversion techniques involve huge matrix
computations, and for this reason these methods fail to work for long
input data sets.

The carrier fit algorithm described above combines properties of the
discrete and continuous spectrum methods. First, we have a
carrier frequency $\nu_0$ and its overtones $k\nu_0$, which
constitutes the discrete aspect. If the modulating waveforms $a_k(t)$
and $b_k(t)$ are smooth enough, the resulting spectrum will
consist of $K$ narrow but continuous frequency bands, accounting for
the continuous aspect. 
By combining some good properties of the standard
methods, we can achieve a compromise, where the discreteness of the
midpoints of the narrow bands allow to ``bridge'' long gaps in time
domain, and the continuity of the local spectra around the mid-frequencies
allows to describe smooth time-dependent changes. 
Our method can essentially be regarded as a quite straightforward
combination of the two classical approaches.

\subsection{Local frequency analysis}

There is another set of methods popular among variable star
researchers. These methods are based on a division of the input data
into separate segments, and then using some standard method to analyse
them. The segmentation can be based on natural fragmentation of the
input data or by formal division of the full data span. In some
methods the segments can be overlapping in time (see recent detailed
implementation by Lehtinen et al~\cite{Lehtinen}). It is assumed that
the changes in the local Fourier spectra or other type statistics computed for
different segments reveal actual trends in frequency structure of the
underlying signal.

A typical example - the $\delta$ Scuti -type star $\theta$ Tuc was
observed during six nights (Stobie \& Shobbrook~\cite{Stobie}), and
data for each night was Fourier analysed seprately.  
The conclusion of the authors was
\begin{quote}
  Photometric observations of $\theta$ Tucanae were Fourier analysed
  for their component frequencies.  No coherent frequency could be
  found spanning the complete data set. It is shown that both the
  frequencies and amplitudes present in $\theta$ Tucanae change on
  time scale as short as 24 hr.
\end{quote}
However, more recent analysis of more extended data sets by other
authors (for example Kurtz~\cite{Kurtz}, Papar\'{o} et
al.~\cite{Paparo}) gave another interpretation to the local variability
of $\theta$ Tuc. It occurred that the varying local spectra result from
constructive and destructive interferences of different coherent
oscillation modes. The interpretation of the local spectra is further
complicated by their low resolution. Above we gave an illustration of
this effect (see Fig.~\ref{Fg08}) in the context of straightforward
local Fourier analysis. This example shows that it is a good practice
to use time-dependent phase diagrams (built by the carrier fit method) and
time-dependent frequency analysis together to get a full understanding
of the underlying process.

Localisation is carried to the limit in the method proposed by
Tsantilas \& Rovithis-Livaniou~(\cite{Tsantilas}). They use a very
similar light curve model to Eq.~(\ref{primi})
\begin{equation}
f(t)=a(t)\sin(b(t)\times t + c(t)),
\end{equation}
where observed data is decomposed into three continuous curves:
amplitude $a(t)$, frequency $b(t)$ and phase $c(t)$. The point
estimates for the three components are then obtained by performing
local non-linear least squares fits. Unfortunately, the presentation
of the light curve in this form is not unique (for instance frequency
changes can be compensated by phase changes) and consequently the
obtained numerical results contain large amount of arbitrariness. In
our method the carrier is fixed and modulations are constrained to be
smooth. This allows to get more useful presentations for the light curves.

\subsection{Time-frequency analysis}

Another set of methods worthwhile to mention in this context 
are different time-frequency transforms. They are all based on a
decomposition of the input data into set of sub-waves. Every sub-wave
(in some methods called wavelet) is localized in time as well as in
frequency. For a large set of this kind of decompositions we refer to
a recent work by Blackman~(\cite{Blackman}) and references therein.
In the context of variable star research, the most important problem with
the standard time-frequency methods is related to the gaps in the input
data. Usually the set of sub-waves is mathematically complete, but
this leads to unrealistic ``filling of gaps''. The sub-waves which are
localized in short time intervals inside observing gaps will get zero
values and as a result of that the structure of gaps will be seen
through in the final two-dimensional spectrum. A very nice
illustration of this point can be found in
Sz\'{a}tmary~(\cite{Szatmary}), where a large catalogue of model wavelet
decompositions is given. 

To overcome problems with gaps it is possible to modify the original
wavelet method by making its resolution to depend on time. This can
be achieved by the use of certain weighting schemes (see
Foster~\cite{Foster} for details) or adaptive wavelets, as done by
Frick et al.~(\cite{Frick}).

In the carrier fit method we use {\it a priori} information to constrain
our sub-waves to be localised in fixed frequency strips around carrier and
possibly around its harmonics. The shortcoming of our formulation is
that our system of sub-waves is not complete. 
Nevertheless, with our algorithm, the gaps can be properly filled in,
which enables to get a continuous phase or frequency diagrams that are
more easy to interpret.

\section{Discussion and conclusions} \label{Discussion}
\subsection{Utility of the carrier fit}

All the simple examples above demonstrate that the carrier fit
procedure itself is not sufficient to perform proper data analysis and
interpretation. It is just a particular method to interpolate
data sets with gaps to obtain a relatively smooth estimate for the
variable star light curve. Interpolation itself has some value only
when the real observed process and its observed values have certain
properties. Most importantly, the process must be governed by some
relatively stationary inherent clocking mechanism whose parameters
change slowly. Fortunately this is the case for many variable stars.

\subsection{Domain of applicability}

By formulation and by algorithmic implementation the proposed utility
can be considered as automatic. However, the obtained results can be
very often misleading. This is why the complete understanding of the
involved ideas is very important. Let us now reiterate major
requirements for successful application of the carrier fit method.

Most importantly, the data to be analysed must come from an object with
a certain internal clocking mechanism, being it rotation,
orbiting or pulsation. This allows to fix, at least approximately, a
proper value for a carrier frequency.

The modulating curves of the basic cycle must be relatively slow so
that the spectral representations of the modulations and carrier frequency
are located in different parts of the spectrum, i.e. the Bedrosian
theorem is valid. It is also important that models for modulations
must be smooth enough to allow ``bridging the gaps'' in observational
sequences.

There must be enough of high quality observational points to allow
adequate interpolation. To avoid over fit and under fit, standard
methods from regression analysis should be used (Draper \&
Smith~\cite{Draper}).

The user of the methods must have a certain expertise to interpret
the obtained phase diagrams. The proper and required intuition for
that can be developed by using model computations with known input
components. As a starting point the examples provided above can be
useful.

Because the method involves inversion of large matrices, 
certain numerical problems may occur. The methods based
on row-wise updating (Givens rotations) to solve the least-squares
computational problems are very useful in this context; for details
see Lawson \& Hanson~(\cite{Lawson}).

The carrier fit method belongs to the class of data analysis
methods called explorative. After getting an intuitively pleasing and
convincing phase diagrams the researcher must quantify the results
using more precise models (with proper estimates of errors etc).

Another important aspect of the new method is the focus on phase
behaviour. For that purpose, the visualization of normalized light
curves is of great importance. 
This facilitates the detection of relevant trends and abrupt changes
in light curves. This is just the aspect which usually remains hidden
when standard multicomponent fit into the data is used.

In this paper we have presented a relatively new but simple class
of data analysis methods which are useful in different astrophysical
contexts. The best way to promote the methods would be their successful
application to actual data sets. 
We are undertaking this effort in the next papers of the series.

The developed software is a part of the general time series package 
(Irregularly Spaced Data Analysis) which is freely available~\footnote{ISDA, http://www.aai.ee/$\sim$ pelt/soft.htm, details by e-mail pelt@aai.ee}.

\begin{acknowledgements}
Ilkka Tuominen sadly passed away on March 19, 2011. We wish to express our 
respect for his importance for the research on magnetically active stars.
Financial support from the Academy of Finland grants 141017 and 218159 is
acknowledged. We also thank the anonymous referee for useful comments on the manuscript.
\end{acknowledgements}


\begin{thebibliography}{}

   \bibitem[2010]{Aerts} Aerts C., Christensen-Dalsgaard J., Kurtz D.W., 2010, Asteroseismology, Springer

   \bibitem[1963]{Barning} Barning F.J.M., 1963, B.A.N., 17, 22

   \bibitem[1963]{Bedrosian} Bedrosian E., 1963, Proceedings of the IEEE, 51, 868

   \bibitem[1998]{Berdyugina98a} Berdyugina S.V., Jankov S., Ilyin I., Tuominen I., Feckel E.C., 1998, A\&A, 334, 863

   \bibitem[2002]{Berdyugina02} Berdyugina S.V., Pelt J., Tuominen I., A\&A, 2002, 394, 505 

   \bibitem[2010]{Blackman} Blackman C., 2010, ApJS, 191, 185

   \bibitem[1998]{Draper} Draper N.R., Smith H, 1998, Applied Regression Analysis, Third edition, Wiley-Interscience
   
   \bibitem[1996]{Foster} Foster G., 1996, AJ, 112, 1709
   
   \bibitem[1997]{Frick} Frick P., Baliunas S.L., Galyagin D., et al, 1997, ApJ, 483, 426

   \bibitem[2008]{Frescura} Frescura F.A.M., Engelbrecht C.A., Frank B.S., 2008, MNRAS, 388, 1693

   \bibitem[1946]{Gabor} Gabor D., 1946, J. Inst. El. Eng, 93, 429

   \bibitem[1973]{Gentleman} Gentleman W.M., 1973, J. Inst. Maths. Applics., 12, 39
 
 
   \bibitem[2002]{Hempelmann} Hempelmann A., A\&A, 388, 540

   \bibitem[1993]{Jetsu93} Jetsu L., Pelt J., Tuominen I., 1993, A\&A, 278, 449
      
   \bibitem[2008]{Jurcsik08} Jurcsik J., S\'odor A., Hurta Zs., et al, 2008, MNRAS, 391, 164

   \bibitem[1990]{Koen} Koen C., 1990, ApJ, 348, 700

   \bibitem[2007]{Korhonen07} Korhonen H., Berdyugina S.V., Hackman T., et al., 2007, A\&A, 476, 881 

   \bibitem[1982]{Kuhn} Kuhn J.R., 1982, AJ, 87, 196
   
   \bibitem[1980]{Kurtz} Kurtz D.W., 1980, MNRAS, 193, 61

   \bibitem[1987]{Lawson} Lawson L.C., Hanson R.J., 1987, Solving Least Squares Problems, Society for Industrial Mathematics

   \bibitem[2011]{Lehtinen} Lehtinen J., Jetsu L., Hackman T., Kajatkari P., Henry G.W., 2011, A\&A, 527, A136

   \bibitem[2010]{Nygren} N\`ygren T., Ulich Th., 2010, Ann. Geophys., 28, 1409

   \bibitem[1996]{Paparo} Papar\'{o} M., Sterken C., Spoon H.W.W., Birch P.V., 1996, A\&A, 315, 400  

   \bibitem[2010]{Poularikas} Poularikas A.D. (editor) 2010, Transforms and Applications Handbook, 3rd edn (CRC Press) 

   \bibitem[2000]{Rodono} Rodon\`o, M., Messina, S., Lanza, A. F., Cutispoto, G., Teriaca, L., 2000, A\&A, 358, 624

   \bibitem[2003]{SC} Schwarzenberg-Czerny, A., 2003, ASP Conf. Series, 292, 383

   \bibitem[1978]{Stellingwerf} Stellingwerf R.F., 1978, ApJ, 224, 953

   \bibitem[2005]{Sterken} Sterken C. 2005, in The Light Time Effect in Astrophysics, ed. Sterken C., ASP Conference Series, Vol. 335, 3
  
   \bibitem[1976]{Stobie} Stobie R.S., Shobbrook R.R., 1976, MNRAS, 174, 401 
   
   \bibitem[1981]{Stirling} Stirling W.D., 1981, Applied Statistics, 30, 204 
   
   \bibitem[1994]{Szatmary} Sz\'{a}tmary K., Vink\'{o} J., G\'{a}l J., 1994, A\&AS, 108, 377
   
   \bibitem[2008]{Tsantilas} Tsantilas S., Rovithis-Livaniou H., 2008, Comm. in Astroseismology, 157, 87

   \bibitem[1996]{Vakman} Vakman D. 1996, IEEE Transactions on Signal Processing, 44, 791 

\end{thebibliography}
\end{document}